# Discovery of unconventional charge-spin-intertwined density wave in magnetic kagome metal GdTi$_3$Bi$_4$

Xianghe Han[1,2,#], Hui Chen[1,2,#], Zhongyi Cao[1,2,#], Jingwen Guo[3], Fucong Fei[3], Hengxin Tan[4], Jianfeng Guo[1,2], Yanhao Shi[1,2], Runnong Zhou[1,2], Ruwen Wang[1,2], Zhen Zhao[1,2], Haitao Yang[1,2], Fengqi Song[3], Shiyu Zhu[1,2], Binghai Yan[4], Ziqiang Wang[5], Hong-Jun Gao[1,2] ✉

[1] Beijing National Center for Condensed Matter Physics and Institute of Physics, Chinese Academy of Sciences, Beijing 100190, PR China

[2] School of Physical Sciences, University of Chinese Academy of Sciences, Beijing 100190, PR China

[3] National Laboratory of Solid State Microstructures, School of Physics, School of Materials Science and Intelligent Engineering, Collaborative Innovation Center of Advanced Microstructures, Nanjing University, Nanjing 210093, China

[4] Department of Condensed Matter Physics, Weizmann Institute of Science, Rehovot, Israel

[5] Department of Physics, Boston College, Chestnut Hill, MA 02467, USA

[#]X.H, H.C and Z.C contributed equally to this work

✉Correspondence author, Email address: hjgao@iphy.ac.cn

The symmetry breaking and its interplay among spin, charge, and lattice degrees of freedom is crucial for understanding correlated quantum states such as charge density waves (CDWs) and unconventional superconductivity. Here, we report the discovery by low-temperature scanning tunneling microscopy/spectroscopy of unconventional charge-spin-intertwined density waves in magnetic kagome metal GdTi$_3$Bi$_4$, which exhibits the one-third magnetization plateau. We reveal the emergence of 3Q CDWs incommensurate with the crystalline lattice in both periodicity and orientation, breaking all mirror and rotation symmetries. The CDW exhibits incommensurate-commensurate transitions in an applied magnetic field and transitions between 3Q and 1Q CDWs as a function of field and temperature, accompanied by changes in the spatial symmetries. Remarkably, the quantum and classic melting of the CDWs exhibits a phase structure which is consistent with the magnetization phase diagram of bulk GdTi$_3$Bi$_4$, providing strong evidence for the intertwined charge-spin density wave order. The origin of the charge-spin intertwinement is further evidenced by the observed hybridization between itinerant electrons and Gd local moments. Our findings uncover an unconventional form of charge-spin orders and offer new insights into a broad class of multi-components density wave formation in kagome and other correlated quantum materials.

The Symmetry breaking[1–4] and its coupling with various degrees of freedom[5,6] play a fundamental role in the emergence of correlated quantum states. Among them, charge density waves (CDWs), which represent electronic states that disrupt lattice translational symmetry, have been a key subject of study in condensed matter physics[5,7,8]. Recently, CDWs have been recognized as essential elements in the complex electronic states of quantum materials, where they interact with phenomena such as superconductivity[9–13], Hall effect[14–16] and non-equilibrium phases[17–19]. To fully understand these complex states, it is essential to experimentally probe the coupling of CDWs with various degrees of freedom phases and identify the associated local symmetry breakings. As the CDWs inherently break the translational symmetry of the underlying lattice, additional symmetry breaking would give rise to novel electronic phenomena such as nematicity[11,20–22], chirality[23–26] and odd-parity state[27]. Moreover, the coupling of charge orders with other parameters such as orbital textures[28,29] and Cooper pair modulations[30,31] can generate exotic intertwined orders including pair density waves[32]. Despite these insights, the experimental exploration of CDW intertwinement with spin textures in correlated quantum materials remains rarely reported.

In magnetic metals, the spin and charge are expected to be intricately intertwined. Spin density waves which are crucial for understanding the formation of unconventional superconductivity[33–35], have been theoretically predicted in the magnetic metals[36,37]. For instance, in the antiferromagnetic metal FeGe, a CDW order is found to coexist with magnetic order, but remains separate from the antiferromagnetism[38,39]. Recently, kagome magnets incorporating various rare-earth elements (RE) elements have been shown to host diverse magnetic orders alongside exotic quantum transport behaviors. In $REMn_6Sn_6$, the coexistence of magnetism stemming from the Mn kagome lattice and the RE atoms results in a spectrum of magnetic ground states[40–42]. In contrast, magnetic metals such as $REV_6Sn_6$[43–45], $REV_3Sb_5$[46,47] and $RETi_3Bi_4$[48–51], feature nonmagnetic V/Ti-kagome lattices, where the RE atoms dictate the magnetic behavior, giving rise to various magnetic states. Among these, $RETi_3Bi_4$ exhibiting unique fractional magnetizations and energy band[52–56], stand out as specific platform for the exploring the interplay of magnetic ordering from RE layers with density wave formation arising from nonmagnetic kagome layers, with the potential for unconventional charge-spin coupling.

In this work, we report the discovery of charge-spin-intertwined density waves in the rare-earth kagome metal $GdTi_3Bi_4$, which exhibits an antiferromagnetic (AFM) ground state and a one-third plateau magnetization arising from one-dimensional Gd zigzag chains. Using scanning tunneling microscopy/spectroscopy (STM/STS), we reveal the presence of three-components (3Q) charge orders at the Gd-terminated surface. In the AFM ground state, the charge order is incommensurate with the crystalline lattice in both length and directions, breaking all mirror and rotation symmetries. Under an applied magnetic field, the CDW undergoes a transition from an incommensurate to a commensurate order, along with field- and temperature-driven shifts between 3Q and 1Q configurations, accompanied by symmetry changes. As the magnetic field increases from ground state to the 1/3 plateau state, the restored rotation and mirror symmetries results in a nearly commensurate $3a_0 \times 3a_0$ charge order. As temperature increases, the 3Q charge order melts into a 1Q charge order, which can revert to the 3Q configuration upon the application of a magnetic field. Notably, the quantum and classic melting of the CDWs exhibits a phase structure that closely aligns with the magnetization phase diagram of bulk $GdTi_3Bi_4$, offering strong evidence of charge-spin intertwinement. This intertwinement is further supported by the observed hybridization between itinerant electrons and localized Gd moments.

$GdTi_3Bi_4$ has a layered orthorhombic lattice structure in the space group Fmmm (Fig. 1**a**), The prototype structure consists of four slightly distorted Ti kagome-lattice layers per unit cell, which is larger than that of kagome superconductor $CsV_3Sb_5$[48]. Another interesting structural motif of interest is the formation of Gd-Gd zigzag chains along *a* direction (Fig. 1**a**). The weak interlayer interactions between adjacent Gd/Bi layers (light green shade in Fig. 1**a**) results in the Gd terminated surface (Fig. 1**b**) after cleaving the $GdTi_3Bi_4$ crystal[51]. The $GdTi_3Bi_4$ crystal exhibits quasi-two-dimensional energy band. The Fermi surface contains three pairs of bands (Fig. 1**c**): triangular bands from Ti 3*d* orbitals (blue), quasi-1D band from Gd 5*d* orbitals (purple) and an inner band from Bi 6*p* orbitals (orange). The hybridization of the Gd-Gd zigzag chain with Bi 6*p* orbital band results in the elongated shape of the inner pocket. The uniaxial direction of the ellipse is perpendicular to quasi-1D band.

The $GdTi_3Bi_4$ exhibits an antiferromagnetic ground state below a Néel temperature of $T_N \sim 14$ K, as demonstrated by magnetization measurements at a temperature range from 2 K to 100 K (Fig. 1**d**). In addition, $GdTi_3Bi_4$ shows a specific 1/3 plateau state with the magnetic field along the *c*-axis between $H_{c1}$ and $H_{c2}$. When the field is higher than $H_{c2}$, it turns to fully polarized ferromagnetic states (Fig. 1**e**).

The electronic states of GTB at atomic scale are also studied by performing low-temperature STM/S. The STM image of Gd terminated surfaces show flat terrace with randomly distributed defects (Fig. 1**f**). The height difference between two adjacent terraces is about 1.3 nm, consistent with the lattice constant of 1.3 nm along *c* direction (Fig. S1). The typical atomically-resolved STM images at relatively large sample bias, e.g. -100 mV, show hexagonal lattice with Gd atoms as Gd layer is slightly higher than the Bi atomic layer (Fig. 1**f**). When approaching the tip closer to the sample surface by using small sample bias *e.g.* -10 mV, the STM images resolve both Gd and Bi atoms (Fig. 1**g**, details please see Fig. S2). The local density of states (LDOS) at Gd terminated surface display a V-shape metallic feature around the Fermi level ($E_F$) with a broad conductance peak at 270 mV, as evidenced by the spatially averaged d$I$/d$V$ spectra (Fig. 1**h**). The broad peak is spatially localized at the Gd atoms, as evidenced by the spatial distribution of d$I$/d$V$ spectra (Fig. S3). In addition, the broad peak split with increasing external magnetic field perpendicular to the sample surface (Fig. 1**h**). Thus, we attribute the broad peak to a localized magnetic state from Gd zigzag chains.

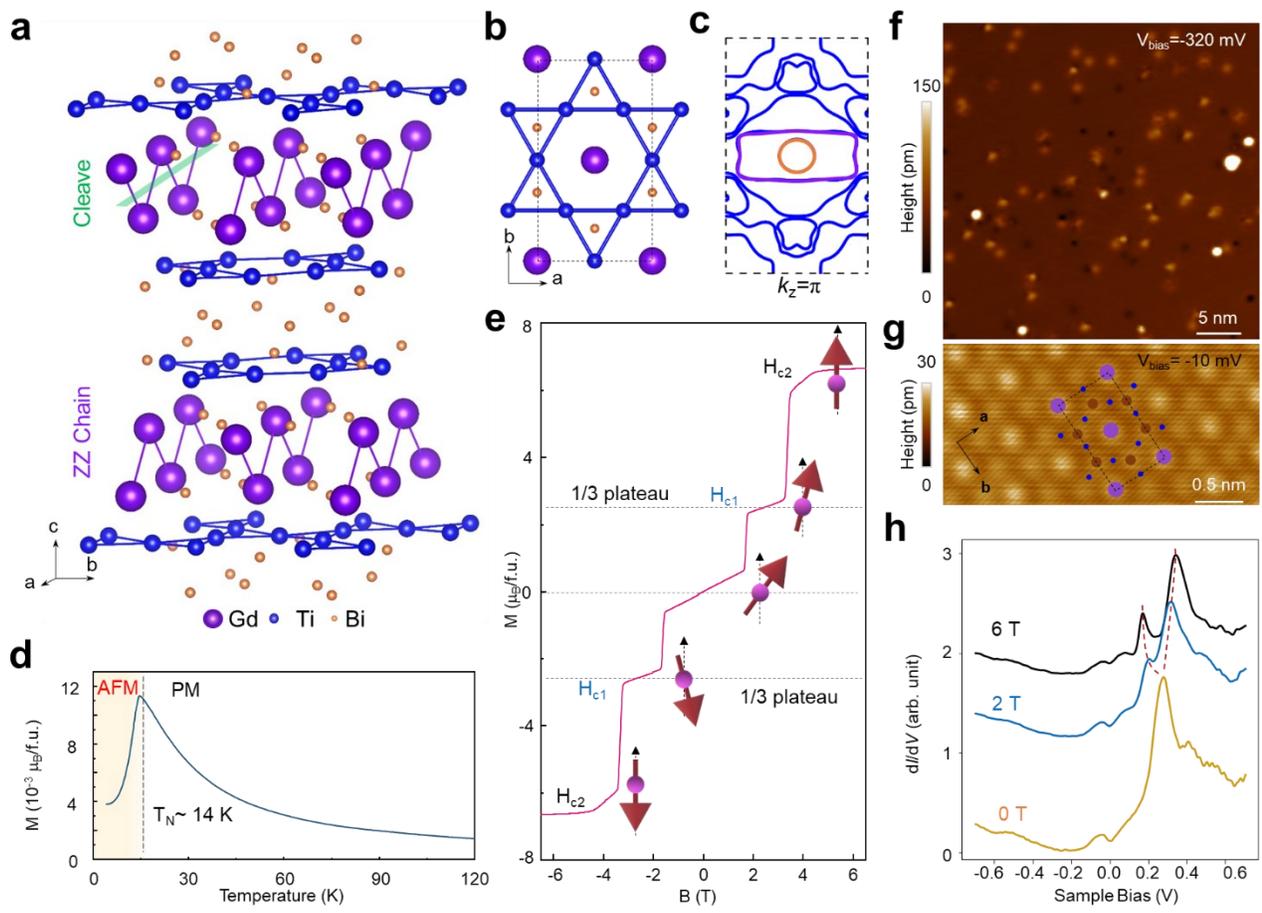

**Fig. 1 Structural, electronic and magnetic properties of the GdTi$_3$Bi$_4$ crystal. a,** Atomic model of GdTi$_3$Bi$_4$ crystal, showing the bilayer kagome layer and Gd zigzag chain. Green shade indicates the cleaving plane. **b,** Top view of the atomic model showing as-cleaved Gd surface with underlying kagome-lattice TiBi layer. **c,** Calculated Fermi surface, showing contributions from the energy band of Gd zigzag chain (purple) and Ti kagome layer (blue). **d,** Temperature dependence of magnetization under a fixed magnetic field of 100 Oe, showing the antiferromagnetic transition occurs at $T_N$ ~14 K. **e,** Field dependence of magnetization, showing the two steps of magnetic transition with $H_{c1}$~ 1.3 T and $H_{c2}$~ 3.2 T. **f,g** Relatively large-scale (**f**) and atomically resolved (**g**) STM images showing the Gd terminated surfaces with a quasi-hexagonal lattice. The atomic model is overlaid for clarification of Gd and Bi atoms. Tunneling parameter: (**f**), $V_{set}$ = -320 mV, $I_{set}$ = 450 pA (**g**), $V_{set}$ = -10 mV, $I_{set}$ = 100 pA. **h,** The magnetic dependent d$I$/d$V$ spectrum on Gd atom, showing a magnetic state at about 0.27 eV above $E_F$. $V_{set}$ = 700 mV, $I_{set}$ = 1000 pA, $V_{mod}$ = 2.5 mV.

In addition to the local magnetic state at 270 meV, there are some electronic states at low energies in the d$I$/d$V$ spectra of Gd surface. The spatial distribution of LDOS around $E_F$ e.g. 20 meV (Fig. 2**a**), shows the real-space LDOS oscillation around the defects originated from quasiparticle interferences (QPI). These QPI signals result in an elongated circular pattern centered at $\Gamma$ point in the corresponding Fourier transform (FT) image of d$I$/d$V$ map (Fig. 2**b**). The elongated circular QPI is originated from the intra-band scattering of ellipse band. The long axis of such QPI pattern corresponds to the uniaxial direction of the ellipse band from Bi band. By comparing the chain direction with the calculated Fermi surface, we attribute the long axis of elongated circular QPI pattern ($Q_B^3$ in Fig. 2**b**) to the zigzag chain direction (*a* direction).

The FT cut along one direction (orange dashed line in Fig. 2**b**) in the *q* space shows a dispersing parabolic curve (dashed dotted curve in Fig. 2**c**), corresponding to the expending elongated circular QPI patterns which grows outwards with increasing energy. The dispersive parabolic band is interrupted by the localized magnetic state at an energy of 270 meV. Such interruption is attributed to the hybridization between itinerant electrons and local moments from Gd zigzag chains[57,58].

We then study the itinerant electrons states near $E_F$. In the FT of d$I$/d$V$ maps at low energies, there are three pairs of additional peaks around the circular QPI pattern (outlined by blue, green and red squares in Fig. 2**b**). To understand the origin of these extra peaks in the FT, we collect the d$I$/d$V$ maps at various energies, i.e. $g(r, E)$ (*r* is the real space location while *E* is the energies). The $g(r, E)$ maps in large energies (*e.g.* -280 meV, Fig. 2**d**) show super-modulations propagating along three directions, forming a superlattice over the pristine structural lattice of Gd atoms (inset of Fig. 2**d**). These additional modulation gives rise to the additional peak structures seen in the FTs, termed as $q^i$ (i=1, 2, 3) vectors (Fig. 2**e**).

To distinguish signals between dispersive QPI and non-dispersive CDW orders, we study the energy dependence of the $q^i$ vectors associated with the peaks in the FT. To do this, we obtain linecuts from FTs of the LDOS maps along the three directions in momentum space and plot this as a function of energy (Fig. 2**f**). Contrary to energy-dispersive features such as QPIs, we find that the magnitudes of these three *q* vectors show no energy dependence (details see Fig. S4). This indicates that the observed modulations arise from a CDW order in this material. We label the CDW vectors $q_{CDW}^i$ by the blue ($q_1$), red ($q_2$) and green ($q_3$) squares. The periodicities of CDWs are close to but not equal to three times of the lattice constant $a_0$ (the vector length of $|q_{CDW}^1|=0.30|Q_B^1|$, $|q_{CDW}^2|=0.37|Q_B^2|$ and $|q_{CDW}^3|=0.40|Q_B^3|$). In addition, there is a small misorientation between $q_{CDW}^i$ and $Q_B^i$, indicating that all three CDWs are incommensurate with the underlying lattice (Fig. 2e).

The three vectors do not ideally align with crystalline lattice (cyan lines in Fig. 2e), resulting in distorted hexagonal pattern (Fig. 3a). The rotation and all mirror symmetries of the distorted hexagonal pattern are breaking, indicates that the three CDW orders have independent order parameters. In addition, we filter the FT images by passing $q_{CDW}^i$ and $Q_B^i$ spots and produce the inverse FT maps back to the real space (Fig. 3b), highlighting the CDWs are incommensurate with both the periodicity and orientation of the underlying lattice.

The possible hybridization between itinerant states and localized magnetic states immediately leads to the question of whether the CDW vectors varys with magnetic states. Thus, we study the evolution of three vectors of CDW with the magnetic field perpendicular to the sample surface ($B_z$) at 0.4 K. In the AFM state ($B_z<\mu H_1$), the 3Q CDW with distorted hexagon shape remain unchanged (Fig. 3c). While in the 1/3 plateau state by increasing $B_z > \mu H_1$, the distorted hexagon shape connecting three pairs of CDW peaks become more symmetric. Further increasing $B_z$ close to $\mu H_2$, the 3Q CDWs gradually melt and are eventually invisible in the FM state when $B_z > \mu H_2$ (Fig. 3c). The CDWs disappear at the fully polarized ferromagnetic state, indicating that the charge orders are locked to the antiferromagnetic spin orders.

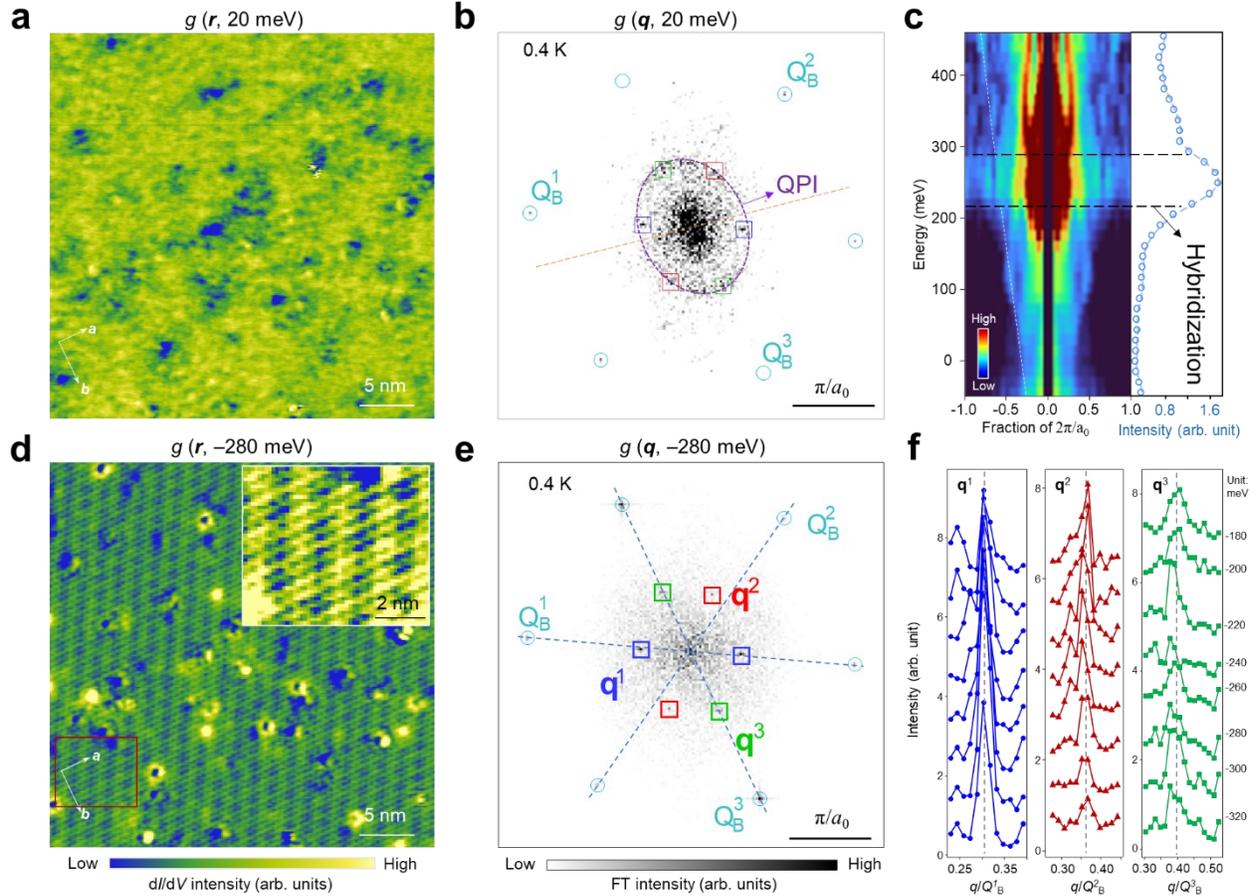

**Fig.2. Spectroscopic imaging of the quasiparticle interference pattern and three distinct density wave orders. a,b** The d$I$/d$V$ map at 20 meV, g($r$, 20 meV) (**a**) and corresponding FT, g($q$, -20 meV) (**b**), showing dominant circular QPI patterns around impurities. $V_{set}$ = 20 mV, $I_{set}$ = 200 pA, $V_{mod}$ = 3 mV. **c**, FT cut of g($E$, $q$) along the yellow dotted line in (**b**), showing the dispersions of circular QPI patterns are interrupted by the local magnetic states at 270 meV. Left panel shows the energy dependent FT intensity, indicating the local magnetic states at 270 meV. **d,e** The g($r$, -280 meV) (**d**) and g($q$, -280 meV) (**e**), showing three pairs of density wave peaks ($q^i$). $V_{set}$ = -280 mV, $I_{set}$ = 450 pA, $V_{mod}$ = 7 mV. **f**, Linecuts in Fourier space along the three directions along $q^i$. The positions of $q^i$ do not change in a large energy range from -180 to -320 meV, consistent with the formation of density wave order.

The three $q_{CDW}^i$ vectors satisfy the relationship $q_{CDW}^1+q_{CDW}^2+q_{CDW}^3=0$, independent with the magnetic field (Fig. S7). We term such relationship as a triangle relation, i.e. connecting three vectors result in the formation of a triangle (Figs. 3**a, d**). The symmetric hexagon formed by the three vectors in the 1/3 plateau magnetization state *e.g.* 2.25 T show restored rotation and mirror symmetries, featuring a nearly commensurate $3a_0 \times 3a_0$ charge (Fig. 3**d**). In addition, by filtering the FT images by passing $q_{CDW}^i$ and $Q_B^i$ spots at 1/3 plateau state, *e.g.* 2.25 T and producing the inverse FT maps back to the real space, we observe $3a_0 \times 3a_0$ superlattice nearly commensurate with the underlying crystalline lattice (Fig. 3**e**).

To demonstrate the robustness and repeatability of the field-induced switching of 3Q CDWs between AFM state and 1/3 plateau state, we systematically study the CDWs under the sequence of magnetic field of 0 T-(-2.25 T)-0 T-(2.25 T)- 0 T (Figs. 3**f,g** details see Fig. S6). To quantitively study the variation of 3Q CDWs with external magnetic field, we define the relative angle and relative length by comparison to the Bragg peaks. The relative angle is the angle difference between the CDW vectors and corresponding Bragg vectors. The positive (negative) value means the angle by rotating the CDW vectors towards Bragg vectors in a counterclockwise (clockwise) manner. For the relative angle, the three angles transform from finite angle to zero angle when $B_z$ increase to $\mu H_1$ and vice versa (Fig. 3**f**). Meanwhile, the values of the three relative lengths ($|q_{CDW}^i|/|Q_B^i|$) deviate from 1/3 at 0 T and approach closer to 1/3 when $B_z$ increases to $\mu H_1$ (Fig. 3**g**). Therefore, the incommensurate (IC) CDW in the AFM state evolves into a nearly commensurate (NC) $3a_0 \times 3a_0$ order in the 1/3 plateau state.

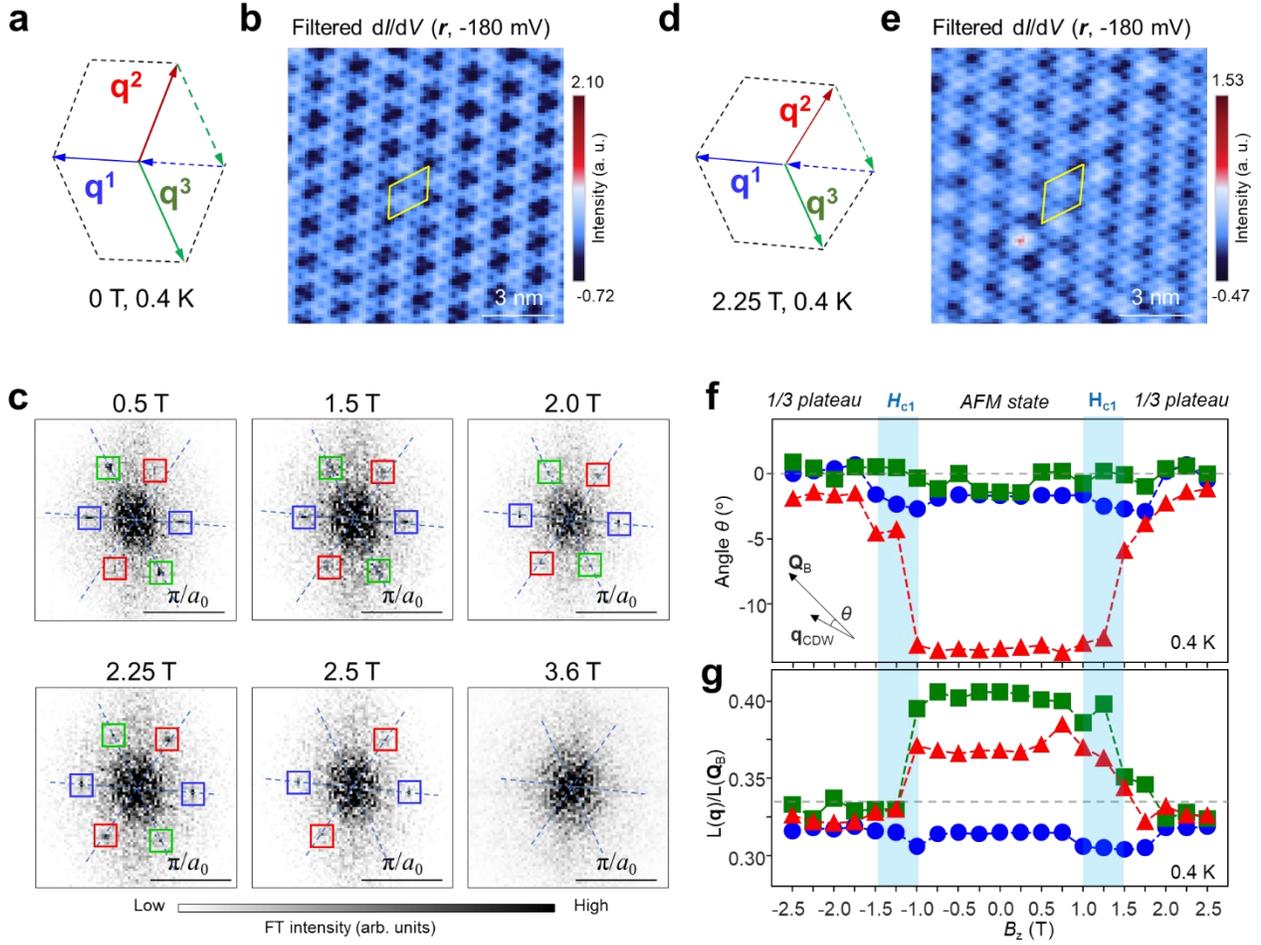

**Fig. 3. Evolution of 3Q CDW with perpendicular magnetic field. a**, Schematic of three density wave vectors $q^i_{CDW}$ at 0 T. Three vectors fit triangular relation, breaking rotation and mirror symmetries in the q space. **b**, The filtered d$I$/d$V$ maps by passing the $q^i_{CDW}$ and $Q_B$ spots in the FT images, showing the incommensurate CDW with crystalline lattice in both periodicity and orientation. **c**, FTs of d$I$/d$V$ maps at a perpendicular magnetic field ($B_z$) of 0.5, 1.5, 2.0, 2.25, 2.5 and 3.6 T, respectively, showing that three CDW vectors changes with $B_z$. The dashed blue lines highlight the orientation differences between Bragg peaks ($Q_B$) and density wave peaks ($q^i$). The CDWs are fully melted at 3.6 T. **d**, Schematic of three density wave vectors $q^i_{CDW}$ at 2.25 T. Three vectors fit triangular relation, restoring rotation and mirror symmetries in the q space. **e**, The filtered d$I$/d$V$ maps after passing the $q^i_{CDW}$ and $Q_B$ spots in the FT images at 2.25 T, showing a nearly commensurate $3a_0 \times 3a_0$ CDW. **f,g** plots of orientation angle differences (**f**) and relative length (**g**) with increasing $B_z$, showing the field induced incommensurate-commensurate transition of CDW order. The background color indicates the critical field $H_{c1}$. For all d$I$/d$V$ data in this Figure, $V_{set}$ = -180 mV, $I_{set}$ = 450 pA, $V_{mod}$ = 8 mV.

Except for the quantum melting induced by the magnetic field[59], increasing temperature is expected to induce the thermal melting[60] of the 3Q CDW states in GTB as well. We measure and compare the FTs of d$I$/d$V$ maps at various sample temperature over an identical area of the sample, acquired under the same experimental conditions (Fig. 4**a**). When the temperature increases from 0.4 K to 4.2 K, two of the three CDW (**q**$^2$ and **q**$^3$) peaks disappear, resulting in the formation of unidirectional CDW order with vector **q**$^1$ (Fig. 4**a** and Fig. S8). Further increasing the temperature close to the Néel temperature ($T_N$~14 K), the unidirectional **q**$^1$ are nearly suppressed. The $T_{CDW}$ ~$T_N$ indicates that the CDW orders are strongly correlated to the AFM orders, which is distinct from the CDWs in kagome antiferromagnet FeGe where $T_{CDW}$ is much lower than $T_N$.[38]

The transition temperature from 3Q-CDW to 1Q-CDW occurs at a temperature range of 2-4 K. To quantitatively evaluate the melting of the CDW phase, we extract FT linecuts from the center of the FT through each of the three CDW peaks as a function of temperature from 1.0 K to 2.6 K (details see Fig. S6). By plotting the peak intensities as a function of temperature, we find that the $q^1$ peak keeps unchanged while the other two peaks get progressively suppressed with increasing temperature (Fig. S9). The critical temperature for $q^2$ is 2.5 K, which is higher than that of $q^3$ (2.2 K). The distinct critical temperature for the suppression of three CDW peaks further demonstrates the independent order parameters of the 3Q CDW orders. The critical temperature around 2.2 K corresponds to the magnetic state transition in the MFM measurements (Figs. S10 and S11), further supporting the strong correlation between CDWs and magnetic orders.

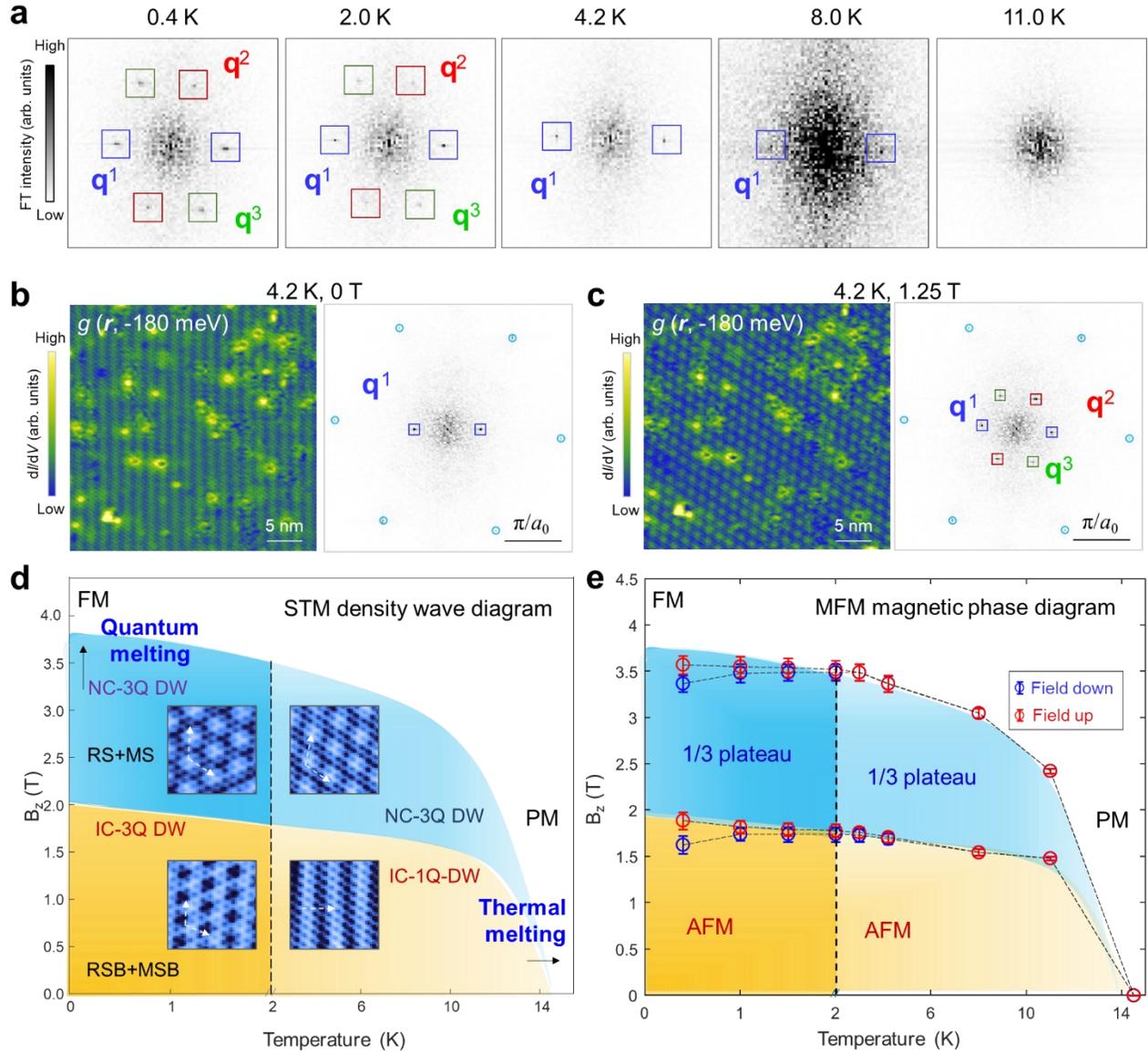

**Fig. 4. Temperature evolution and field-temperature diagram of spin-charge-intertwined density wave in GdTi$_3$Bi$_4$. a,** FTs of d$I$/d$V$ maps at a temperature of 0.4 K, 2.0 K, 4.2 K, 8.0 K and 11.0 K, respectively, showing the 3Q-1Q transition across 2 K and thermal melting across $T_N$~14 K. **b,c** The d$I$/d$V$ map at -200 mV and corresponding FT at a temperature of 4 K and a $B_z$ of 0 T (**b**) and 1.25 T (**c**), showing the transition from 1Q density wave at 0 T to 3Q density waves at 1.25 T. Tunneling parameter: (**b**), $V_{set}$ = -150 mV, $I_{set}$ = 450 pA, $V_{mod}$ = 8 mV. (**c**), $V_{set}$ = -150 mV, $I_{set}$ = 500 pA, $V_{mod}$ = 8 mV. **d,e** Approximate temperature vs magnetic field schematic phase diagram of density waves from STM measurements (**d**), which shows strong correlation with the magnetic phase diagram based on the MFM measurements (**e**). RS(B): rotation symmetry breaking; MS(B): mirror symmetry breaking; IC: incommensurate; NC: nearly commensurate; DW: density wave; AFM: antiferromagnetic state; FM: ferromagnetic state; PM: paramagnetic state;

We also study the evolution of CDWs with external magnetic field at 4.2 K. When the field increases to above $\mu H_1$, the 1Q CDW turn back to 3Q order (Fig. 4**b**, **c**). The 3Q orders gradually suppress when the magnetic field increases to $\mu H_2$. (Fig. S12).

Finally, we construct a phase diagram of CDW with temperature and magnetic field. The transition of density waves (Fig. 4**d**) is consistent with the magnetic transition phase diagram from magnetization and MFM (Fig. 4**e**, details see method part), demonstrating the strong intertwining between spin and charge density waves. The field mainly induces the incommensurate-commensurate transition together with restored symmetries. The fully polarized FM state results in the quantum melting of density waves. On the other hand, warming temperature results in the 3Q-1Q transition. The PM state leads to the thermal melting of density waves.

In conclusion, we have, for the first time, observed the formation of an incommensurate 3Q CDWs breaking all rotation and mirror symmetries in the magnetic crystal $GdTi_3Bi_4$. The incommensurability is tunable through the application of an external magnetic field, and these phenomena occur concurrently with the critical fields of antiferromagnetic and 1/3 plateau magnetization phases. Furthermore, as the temperature increases, a transition from 3Q to 1Q occurs around 2 K, followed by thermal melting across the Néel temperature. This behavior, underscores the observed strong coupling between the itinerant electrons and the local Gd moments. the quantum and classic melting of the CDWs exhibits a phase structure which is consistent with the magnetization phase diagram of bulk $GdTi_3Bi_4$, providing strong evidence for the intertwined charge-spin density wave order. By uncovering a charge-spin intertwined order, our results establish a new paradigm for manipulating spin-charge intertwined density waves via external magnetic fields, offering potential applications in spintronics and next-generation topological materials engineering.

## Methods

**Single crystal growth of GdTi$_3$Bi$_4$.** The flux method is utilized to synthesize the high–quality single crystals. The starting elements of Gd: Ti: Bi in the molar ratio of 1.2: 1: 20 are mixed in an alumina crucible and sealed in a quartz ampule. The ampule is placed in a furnace and heated to 1000 °C at a rate of 1 °C/min. After maintaining it at 1000 °C for 16 h, the ampule is slowly cooled down to 520 °C in 2000 min. The hexagonal shaped single crystals are obtained after centrifuging in order to remove the excess flux.[51]

**Scanning tunneling microscopy/spectroscopy.** The samples used in the STM/S experiments were cleaved at low temperature (80 K) and immediately transferred to an STM chamber and cooled down to 4.2 K. Experiments were performed in an ultrahigh vacuum (1×10$^{-10}$ mbar) ultra-low temperature STM system equipped with external magnetic field perpendicular to the sample surface. The lowest base temperature is 0.4 K with an electronic temperature of 650 mK (calibrated using a standard superconductor, Nb crystal). The magnetic field up to 11 T was applied using the zero-field cooling technique. All the scanning parameters (setpoint voltage $V_s$ and tunneling current $I_t$) of the STM topographic images are listed in the figure captions. The d$I$/d$V$ spectra were acquired by a standard lock-in amplifier at a modulation frequency of 973.1 Hz, the modulation bias ($V_{mod}$) is listed in the figure captions. Non-magnetic tungsten tips were fabricated via electrochemical etching and calibrated on a clean Au(111) surface prepared by repeated cycles of sputtering with argon ions and annealing at 500 °C. To remove the effects of small piezoelectric and thermal drifts during the acquisition of d$I$/d$V$ maps, we apply the Lawler-Fujita drift-correction algorithm[61] in angle and relative length ($|q_{CDW}^i|/|q_B^i|$) calculation in Fig. 3, which corrects the atomic Bragg peaks in the FTs of d$I$/d$V$ maps to a single pixel to enhance calculational accuracy. STM images to be exactly equal in magnitude and 60° apart.

**Magnetic Force Microscopy.** The MFM experiments were captured by a commercial magnetic force microscope (attoAFM, attocube) using a commercial magnetic tip (MFMR, Nanoword) based on a closed-cycle He cryostat (Bluefors) and dilution refrigerator system (Bluefors), combined with a vector magnet (9-3-1, Bluefors). The scanning probe system was operated at the resonance frequency of the magnetic tip, approximately 70 kHz. The MFM images were taken in constant height mode with the scanning plane nominally ~250 nm above the sample surface. The MFM signal, i.e., the resonance frequency shift ($\Delta f$), is proportional to the out-of-plane stray field gradient. Darker (brighter) magnetic contrast indicated the more attractive (repulsive) interaction between the magnetic tip and the sample stray field. GdTi$_3$Bi$_4$ single crystals were cleaved under ambient conditions to expose fresh surfaces before conducting MFM measurements.

**Magnetic transition at 2 K.** We investigate the magnetic transition below 2 K by performing magnetic force microscopy (MFM) measurements of GdTi$_3$Bi$_4$ single crystals at temperatures. At 0.4 K, we observe the formation and evolution of uniformly stripe domains along the *a*-axis direction during the magnetic transition process due to the quasi-1D magnetic structure of GdTi$_3$Bi$_4$, leading to a pronounced in-plane magnetic anisotropy (Fig. S10). During the increasing magnetic field procedure, stripe domains emerge at 1.79 T and vanish completely at 1.99 T, indicating a magnetic transition within a range of 0.20 T. However, during the decreasing magnetic field procedure, the stripe domains reappear between 1.53 T and 1.73 T, suggesting a magnetic field hysteresis of 0.24 T (Figs. S10,11). In contrast, as the temperature increases to 4 K, the stripe domains on longer align strictly along the *a*-axis and exhibit some deviation (Fig. S11). Meanwhile, the hysteresis becomes negligible, and the magnetic transition window narrows. The switching between the presence and absence of hysteresis behavior, occurring at temperatures below $T_N$, indicates a transition in the microscopic spin structure[62–65]. Therefore, the difference in the orientation of stripe domains and the disappearance of hysteresis suggests the presence of two distinct magnetic structures, with a phase transition occurring between 0.4 K and 4.0 K. The precise temperature-dependent measurements reveal the vanishing of hysteresis around 2.5 K, indicating the manifestation of a spin phase transition, while the second magnetic transition supports the same inference. The spin structure is expected to show more pronounced characteristics at the atomic scale, and potential coupling with electronic states may also be observed. Accordingly, we plot H-T phase diagram of GdTi$_3$Bi$_4$ by extracting the magnetic transition window and field hysteresis at various temperatures (Fig. 4**e**).

**Repeatable observation of CDWs in GdTi$_3$Bi$_4$.** Except for the data shown in the main Figures, we have verified the existence of the CDWs exhibiting the same periodicity and orientation with the crystalline lattice across different samples and tips from three different growth batches (Fig. S8).

**Author Contributions:** H.-J. G. and H. C. design the experiments. X.H., H.C., Z.C. and Y.S. performed the STM/S experiments and data. J.-W.G., Y.F., F.S., Z.Z., R.W. and H.Y. prepared the GdTi$_3$Bi$_4$ crystals and performed the transport experiments. J.-F.G., R.Z. and S.Z. perform the MFM measurements. H.T. and B.Y. did the DFT calculations. Z.W. did the theoretical consideration. X. H., H. C., Z.W. and H.-J.G. wrote the manuscript with input from all other authors. H.-J.G. supervised the project.

**Competing Interests:** The authors declare that they have no competing interests.

**Data availability**

All data that support the findings of this study are present in the paper and the Supplementary Information. Further information can be acquired from the corresponding authors upon request.

# References

1. Fischer, M. H., Sigrist, M., Agterberg, D. F. & Yanase, Y. Superconductivity and Local Inversion-Symmetry Breaking. *Annu. Rev. Condens. Matter Phys.* **14**, 153–172 (2023).
2. Du, L. *et al.* Engineering symmetry breaking in 2D layered materials. *Nat. Rev. Phys.* **3**, 193–206 (2021).
3. Ghosh, S. K. *et al.* Recent progress on superconductors with time-reversal symmetry breaking. *J. Phys. Condens. Matter* **33**, 033001 (2020).
4. González-Cuadra, D., Bermudez, A., Grzybowski, P. R., Lewenstein, M. & Dauphin, A. Intertwined topological phases induced by emergent symmetry protection. *Nat. Commun.* **10**, 2694 (2019).
5. Fradkin, E., Kivelson, S. A. & Tranquada, J. M. Colloquium: Theory of intertwined orders in high temperature superconductors. *Rev. Mod. Phys.* **87**, 457–482 (2015).
6. Fernandes, R. M., Orth, P. P. & Schmalian, J. Intertwined Vestigial Order in Quantum Materials: Nematicity and Beyond. *Annu. Rev. Condens. Matter Phys.* **10**, 133–154 (2019).
7. Chen, C.-W., Choe, J. & Morosan, E. Charge density waves in strongly correlated electron systems. *Rep. Prog. Phys.* **79**, 084505 (2016).
8. Baggioli, M. & Goutéraux, B. Colloquium: Hydrodynamics and holography of charge density wave phases. *Rev. Mod. Phys.* **95**, 011001 (2023).
9. Gerber, S. *et al.* Three-dimensional charge density wave order in $YBa_2Cu_3O_{6.67}$ at high magnetic fields. *Science* **350**, 949–952 (2015).
10. Wandel, S. *et al.* Enhanced charge density wave coherence in a light-quenched, high-temperature superconductor. *Science* **376**, 860–864 (2022).
11. Joe, Y. I. *et al.* Emergence of charge density wave domain walls above the superconducting dome in $1T$-$TiSe_2$. *Nat. Phys.* **10**, 421–425 (2014).
12. Yu, F. *et al.* Pressure-Induced Dimensional Crossover in a Kagome Superconductor. *Phys. Rev. Lett.* **128**, 077001 (2022).
13. Zheng, L. *et al.* Emergent charge order in pressurized kagome superconductor $CsV_3Sb_5$. *Nature* **611**, 682–687 (2022).
14. Qin, F. *et al.* Theory for the Charge-Density-Wave Mechanism of 3D Quantum Hall Effect. *Phys. Rev. Lett.* **125**, 206601 (2020).
15. Yu, F. H. *et al.* Concurrence of anomalous Hall effect and charge density wave in a superconducting topological kagome metal. *Phys. Rev. B* **104**, L041103 (2021).
16. Golovanova, D., Tan, H., Holder, T. & Yan, B. Tuning the intrinsic spin Hall effect by charge density wave order in topological kagome metals. *Phys. Rev. B* **108**, 205203 (2023).
17. Kogar, A. *et al.* Light-induced charge density wave in $LaTe_3$. *Nat. Phys.* **16**, 159–163 (2020).
18. Wang, Y. *et al.* Axial Higgs mode detected by quantum pathway interference in $RTe_3$. *Nature* **606**, 896–901 (2022).
19. Zhou, F. *et al.* Nonequilibrium dynamics of spontaneous symmetry breaking into a hidden state of charge-density wave. *Nat. Commun.* **12**, 566 (2021).
20. Wu, P. *et al.* Unidirectional electron–phonon coupling in the nematic state of a kagome superconductor. *Nat. Phys.* (2023) doi:10.1038/s41567-023-02031-5.
21. Nie, L. *et al.* Charge-density-wave-driven electronic nematicity in a kagome superconductor. *Nature* **604**, 59–64 (2022).
22. Zhang, Z. *et al.* Nematicity and Charge Order in Superoxygenated $La_{2-x}Sr_xCuO_{4+y}$. *Phys. Rev. Lett.* **121**, 067602 (2018).

23. Ishioka, J. *et al.* Chiral Charge-Density Waves. *Phys. Rev. Lett.* **105**, 176401 (2010).
24. Song, X. *et al.* Atomic-scale visualization of chiral charge density wave superlattices and their reversible switching. *Nat. Commun.* **13**, 1843 (2022).
25. Jiang, Y.-X. *et al.* Unconventional chiral charge order in kagome superconductor $KV_3Sb_5$. *Nat. Mater.* **20**, 1353–1357 (2021).
26. Xing, Y. *et al.* Optical manipulation of the charge-density-wave state in $RbV_3Sb_5$. *Nature* **631**, 60–66 (2024).
27. Kozii, V. & Fu, L. Odd-Parity Superconductivity in the Vicinity of Inversion Symmetry Breaking in Spin-Orbit-Coupled Systems. *Phys. Rev. Lett.* **115**, 207002 (2015).
28. Ritschel, T. *et al.* Orbital textures and charge density waves in transition metal dichalcogenides. *Nat. Phys.* **11**, 328–331 (2015).
29. Achkar, A. J. *et al.* Orbital symmetry of charge-density-wave order in $La_{1.875}Ba_{0.125}CuO_4$ and $YBa_2Cu_3O_{6.67}$. *Nat. Mater.* **15**, 616–620 (2016).
30. Liu, X., Chong, Y. X., Sharma, R. & Davis, J. C. S. Discovery of a Cooper-pair density wave state in a transition-metal dichalcogenide. *Science* **372**, 1447–1452 (2021).
31. Chen, H. *et al.* Roton pair density wave in a strong-coupling kagome superconductor. *Nature* **599**, 222–228 (2021).
32. Chen, H. & Gao, H.-J. Widespread waves spark superconductor search. *Nature* **618**, 910–912 (2023)
33. Agterberg, D. F. *et al.* The Physics of Pair-Density Waves: Cuprate Superconductors and Beyond. *Annu. Rev. Condens. Matter Phys.* **11**, 231–270 (2020).
34. Allred, J. M. *et al.* Double-Q spin-density wave in iron arsenide superconductors. *Nat. Phys.* **12**, 493–498 (2016).
35. Chen, K. *et al.* Evidence of Spin Density Waves in $La_3Ni_2O_{7-\delta}$. *Phys. Rev. Lett.* **132**, 256503 (2024).
36. Chen, Y. *et al.* Intertwined charge and spin density waves in a topological kagome material. *Phys. Rev. Res.* **6**, L032016 (2024).
37. Yu, S.-L. & Li, J.-X. Chiral superconducting phase and chiral spin-density-wave phase in a Hubbard model on the kagome lattice. *Phys. Rev. B* **85**, 144402 (2012).
38. Teng, X. *et al.* Discovery of charge density wave in a kagome lattice antiferromagnet. *Nature* **609**, 490–495 (2022).
39. Teng, X. *et al.* Magnetism and charge density wave order in kagome FeGe. *Nat. Phys.* **19**, 814–822 (2023).
40. Yin, J.-X. *et al.* Quantum-limit Chern topological magnetism in $TbMn_6Sn_6$. *Nature* **583**, 533–536 (2020).
41. Li, Z. *et al.* Discovery of Topological Magnetic Textures near Room Temperature in Quantum Magnet $TbMn_6Sn_6$. *Adv. Mater.* 2211164 (2023) doi:10.1002/adma.202211164.
42. Li, H. *et al.* Manipulation of Dirac band curvature and momentum-dependent g factor in a kagome magnet. *Nat. Phys.* **18**, 644–649 (2022).
43. Jiang, Y.-X. *et al.* Van Hove annihilation and nematic instability on a kagome lattice. *Nat. Mater.* **23**, 1214–1221 (2024) doi:10.1038/s41563-024-01914-z.
44. Di Sante, D. *et al.* Flat band separation and robust spin Berry curvature in bilayer kagome metals. *Nat. Phys.* **19**, 1135–1142 (2023) doi:10.1038/s41567-023-02053-z.
45. Li, H. *et al.* Spin Berry curvature-enhanced orbital Zeeman effect in a kagome metal. *Nat. Phys.* **20**, 1103–1109 (2024) doi:10.1038/s41567-024-02487-z.

46. Ortiz, B. R. *et al.* YbV$_3$Sb$_4$ and EuV$_3$Sb$_4$ vanadium-based kagome metals with Yb$^{2+}$ and Eu$^{2+}$ zigzag chains. *Phys. Rev. Mater.* **7**, 064201 (2023).
47. Pokharel, G. *et al.* Highly anisotropic magnetism in the vanadium-based kagome metal TbV$_6$Sn$_6$. *Phys. Rev. Mater.* **6**, 104202 (2022).
48. Ortiz, B. R. *et al.* Evolution of Highly Anisotropic Magnetism in the Titanium-Based Kagome Metals LnTi3Bi4 (Ln: La⋯Gd$^{3+}$, Eu$^{2+}$, Yb$^{2+}$). *Chem. Mater.* **35**, 9756–9773 (2023).
49. Ortiz, B. R. *et al.* Intricate Magnetic Landscape in Antiferromagnetic Kagome Metal TbTi$_3$Bi$_4$ and Interplay with Ln$_{2-x}$Ti$_{6+x}$Bi$_9$ (Ln: Tb⋯Lu) Shurikagome Metals. *Chem. Mater.* **36**, 8002–8014 (2024) doi:10.1021/acs.chemmater.4c01449.
50. Chen, L. *et al.* Tunable magnetism in titanium-based kagome metals by rare-earth engineering and high pressure. *Commun. Mater.* **5**, 73 (2024).
51. Guo, J. *et al.* Tunable magnetism and band structure in kagome materials RETi3Bi4 family with weak interlayer interactions. *Sci. Bull.* **69**, 2660–2664 (2024).
52. Cheng, E. *et al.* Striped magnetization plateau and chirality-reversible anomalous Hall effect in a magnetic kagome metal. Preprint at https://doi.org/10.48550/arXiv.2409.01365 (2024).
53. Park, P. *et al.* Spin density wave and van Hove singularity in the kagome metal CeTi$_3$Bi$_4$. Preprint at https://doi.org/10.48550/arXiv.2412.10286 (2024).
54. Cheng, E. *et al.* Spectroscopic origin of giant anomalous Hall effect in an interwoven magnetic kagome metal. Preprint at https://doi.org/10.48550/arXiv.2405.16831 (2024).
55. Zhang, R. *et al.* Observation of Orbital-Selective Band Reconstruction in an Anisotropic Antiferromagnetic Kagome Metal TbTi$_3$Bi$_4$. Preprint at https://doi.org/10.48550/arXiv.2412.16815 (2024).
56. Jiang, Z. *et al.* Topological surface states in quasi-two-dimensional magnetic kagome metal EuTi$_3$Bi$_4$. *Sci. Bull.* (2024) doi:10.1016/j.scib.2024.08.019.
57. Trontl, V. M. *et al.* Interplay of Kondo Physics with Incommensurate Charge Density Waves in CeTe$_3$. Preprint at https://doi.org/10.48550/arXiv.2502.04814 (2025).
58. Gyenis, A. *et al.* Quasi-particle interference of heavy fermions in resonant x-ray scattering. *Sci. Adv.* **2**, e1601086 (2016).
59. Aishwarya, A. *et al.* Melting of the charge density wave by generation of pairs of topological defects in UTe$_2$. *Nat. Phys.* **20**, 964–969 (2024).
60. LaFleur, A. *et al.* Inhomogeneous high temperature melting and decoupling of charge density waves in spin-triplet superconductor UTe$_2$. *Nat. Commun.* **15**, 4456 (2024).
61. Lawler, M. J. *et al.* Intra-unit-cell electronic nematicity of the high-Tc copper-oxide pseudogap states. *Nature* **466**, 347–351 (2010).
62. Matsuura, K. *et al.* Low-temperature hysteresis broadening emerging from domain-wall creep dynamics in a two-phase competing system. *Commun. Mater.* **4**, 71 (2023).
63. Baranov, N. V. *et al.* Magnetic phase transitions, metastable states, and magnetic hysteresis in the antiferromagnetic compounds Fe$_{0.5}$TiS$_{2-y}$Se$_y$. *Phys. Rev. B* **100**, 024430 (2019).
64. Maat, S., Thiele, J.-U. & Fullerton, E. E. Temperature and field hysteresis of the antiferromagnetic-to-ferromagnetic phase transition in epitaxial FeRh films. *Phys. Rev. B* **72**, 214432 (2005).
65. Rikvold, P. A., Brown, G., Miyashita, S., Omand, C. & Nishino, M. Equilibrium, metastability, and hysteresis in a model spin-crossover material with nearest-neighbor antiferromagnetic-like and long-range ferromagnetic-like interactions. *Phys. Rev. B* **93**, 064109 (2016).